\documentclass[]{hdsr}
\graphicspath{{figs/}}
\usepackage{array}
\newcolumntype{R}[1]{>{\raggedleft\let\newline\\\arraybackslash\hspace{0pt}}m{#1}}

\usepackage{multirow}

\usepackage{enumitem}
\usepackage[linesnumbered]{algorithm2e}
\usepackage{textcomp}
\usepackage{thm-restate}
\excludecomment{notdone}

\usepackage{color, colortbl}

\theoremstyle{definition}
\newtheorem{example}{Example}

\NewCommandCopy{\proofqedsymbol}{\qedsymbol}
\AtBeginEnvironment{proof}{\renewcommand{\qedsymbol}{\proofqedsymbol}}

\AtBeginEnvironment{example}{%
  \pushQED{\qed}\renewcommand{\qedsymbol}{\proofqedsymbol}%
}
\AtEndEnvironment{example}{\popQED\endexample}

\usepackage{xspace}

\newcommand{\zz}{\mathbb{Z}}

\allowdisplaybreaks

\usepackage{mathtools}

\begin{document}
\newgeometry{bottom=1.95in}

\volumeheader{Just Accepted}{}{10.1162/99608f92.f1065159}
  \vspace*{.2in}

\begin{center}

  \title{Disclosure Avoidance for the 2020 Census Demographic and Housing Characteristics File}
  \maketitle

  \vspace*{.2in}
  
  \begin{tabular}{cc}
    Ryan Cumings-Menon,\upstairs{\affilone}
    Robert Ashmead,\upstairs{\affiltwo}
    Daniel Kifer,\upstairs{\affilone,\affilthree} \\
    Philip Leclerc,\upstairs{\affilone}
    Matthew Spence,\upstairs{\affilone}
    Pavel Zhuravlev,\upstairs{\affilone} 
    John M. Abowd\upstairs{\affilfour} \\
  \\[0.25ex]
   {\small \upstairs{\affilone} U.S. Census Bureau} 
   {\small \upstairs{\affiltwo} NORC at the University of Chicago, formerly U.S. Census Bureau } \\
   {\small \upstairs{\affilthree} Penn State University }
   {\small \upstairs{\affilfour}      Cornell University, formerly U.S. Census Bureau} \\
  \end{tabular}
   \emails{
    \upstairs{*}The views expressed in this technical paper are those of the authors and not those of the U.S. Census Bureau. The Census Bureau has reviewed this data product to ensure appropriate access, use, and disclosure avoidance protection of the confidential source data (Project No. P-7502798, Disclosure Review Board (DRB) approval number:  CBDRB-FY23-DSEP-001).
    }
  \vspace*{0.3in}
\begin{abstract}
    \textcite{abowd20222020} describe the concepts and methods used by the Disclosure Avoidance System (DAS) to produce formally private output in support of the 2020 Census statistical data product releases, with a particular focus on the DAS implementation that was used to create the 2020 Census Redistricting Data (P.L. 94-171) Summary File. In this paper we describe the updates to the DAS that were required to release the Demographic and Housing Characteristics (DHC) File, which provides more granular tables than other statistical data products, such as the Redistricting Data Summary File. We also describe the final configuration parameters used for the 2020 production DHC DAS implementation, error metrics for these production statistical data products, and plans for future experimental data products that provide confidence intervals for confidential 2020 Census tabulations.
\end{abstract}
\end{center}

\vspace*{0.15in}
\hspace{10pt}
  \small	
  \textbf{\textit{Keywords: }} {Differential Privacy, 2020 Census, TopDown Algorithm, Demographic and Housing Characteristics File}


\section{Media Summary}

One of the more granular statistical data products produced for the 2020 Census is the Demographic and Housing Characteristics (DHC) File. This data product provides tabulations for detailed demographic groups all the way down to census blocks, which are the smallest set of geographic areas used by the Census Bureau. This granularity makes this data product particularly useful to researchers, governmental entities, and businesses.

Like the other 2020 Census statistical data products, the Census Bureau’s Disclosure Avoidance System (DAS) was used on the DHC tabulations to protect the confidentiality of respondents. The DAS tracks the cumulative privacy-loss of respondents due to the publication of the 2020 Census data products using a formally private accounting framework. A variety of implementation and tuning choices were required for the DHC DAS execution to balance multiple trade-offs, including between utility for various DHC use cases, confidentiality of respondents, and the granularity of tabulations. To make these choices, the Census Bureau released two DHC demonstration data products based on 2010 data and collected feedback on the feasibility of potential 2020 DHC use cases from subject-matter experts (SMEs), both internally to the Census Bureau and externally. Final approval of these settings and algorithmic choices was made by the Data Stewardship Executive Policy (DSEP) Committee.

One advantage of this approach relative to the disclosure avoidance methods used in previous decennial censuses is transparency; the final DHC DAS tuning parameter choices, DHC DAS codebase, the intermediate noisy tabulation measurements generated within DHC DAS production execution, and approximate accuracy metrics of the perturbations introduced by the DHC DAS are now publicly available. This transparency also makes it possible for the Census Bureau to formulate confidence intervals that account for the perturbations introduced by the 2020 DAS.

\section{Introduction} \label{sec:intro}

The Census Bureau's 2020 Disclosure Avoidance System (DAS) was used to produce a formally private file in support of the 2020 Census Redistricting Data (P.L. 94-171) Summary File (hereafter redistricting data); the redistricting data file use case is described in more detail by \textcite{abowd20222020}. The DAS takes the confidential Census Edited File (CEF) as input, which is a set of microdata files with components that include data on individuals, housing units, group quarters (GQs), and households; the output of the DAS is called the Microdata Detailed File (MDF). In the interest of a concise exposition, we refer readers to \textcite{abowd20222020} for more detail on the DAS implementation  as well as the concepts in the differential privacy (DP) literature \parencite{dwork2006calibrating} that are used by the DAS. This paper is a companion paper to \textcite{abowd20222020}; we build on this earlier work by providing the updates to the DAS that were required to produce formally private output in support of the Demographic and Housing Characteristics (DHC) File. 

The DHC release is an important source of decennial census data for governments, researchers, and businesses because it provides much more granular counts than other sources, such as the redistricting data. This additional granularity supports different use cases than the redistricting data and also raises some technical issues for executing the DAS; the updates described in this paper were motivated by both of these  factors. The DHC release is split into two data universes, and counts for each of these two universes are found in distinct DAS implementations. First, the \textit{household} universe, or \textit{DHCH}, provides both counts for categories of householders as well as counts for various categories of housing unit and group quarters types. For example, for housing unit types, this includes counts of owner-occupied, vacant, and rented housing units, and, for group quarter types, this includes counts of nursing facilities, college/university student housing, and federal prisons. Second, the \textit{persons} universe, or \textit{DHCP}, provides counts of individuals within population groups. 

In addition to providing the algorithmic changes and production settings used for the 2020 DHC DAS production executions, this paper outlines the process the Census Bureau used to investigate candidate DHC DAS production settings, \textit{i.e.}, the DHC accuracy experiments. These accuracy experiments involved collecting, and attempting to satisfy, accuracy targets for DHC statistical data product use cases with minimal privacy-loss budget (PLB). These accuracy targets were provided by both internal Census subject-matter experts (SMEs) and external SMEs, including participants of academic conferences and feedback provided through a National Academies of Science, Engineering, and Medicine's Committee on National Statistics (CNSTAT) meeting in June of 2022 \parencite{cnstat}. External SMEs based their feedback on two 2020 DHC demonstration data product DAS executions that used 2010 data as input, as is described in more detail in Section \ref{sec:accuracy_experiments}. Accuracy metrics from these experiments, and input from SMEs, were used by the Data Stewardship Executive Policy Committee (DSEP) to set the final 2020 DHC DAS production settings. 

The basic top-down structure used in both the 2020 Redistricting Data (P.L. 94-191) Summary File and the Demographic and Housing Characteristics file was an algorithmic decision primarily motivated by the publication requirements of redistricting, specifically, the necessity to publish data with census blocks as the geographic atom. The executive decision that all DHC tables must be aggregation-consistent with the redistricting tables implied that census blocks remain the geographic atom for DHC. Unlike the redistricting data, which were used to draw voting districts whose boundaries could not be specified in advance, all DHC tables used geographic boundaries that were predetermined in the 2020 Census tabulation geography. This allowed the algorithmic and SMEs to tune the geographic spine used in DHC to improve accuracy with smaller allocations of PLB. These efforts, described in this paper, illustrate the intellectual contribution of the DHC for publishing very high dimensional table sets. Other practitioners can use similar methods.

One significant advantage of the formally private methods used by the DAS is that they are sufficiently transparent to allow for accurate statistical inferences, \textit{i.e.}, confidence intervals and hypothesis tests, of the underlying confidential CEF-based tabulations. This paper also describes areas of ongoing work aimed at leveraging this advantage to provide experimental data products with confidence intervals for CEF-based tabulations.

After describing the DAS at a high level in the next subsection, the remainder of the paper is organized as follows. Section \ref{sec:schemas_and_constrs} describes the schemas and constraints used within DHCH and DHCP DAS implementations. Section \ref{sec:implementation_details} describes updates to the computational environment used for the DAS DHC implementation. Section \ref{sec:accuracy_experiments} describes the process used to set the 2020 production DHC DAS configuration parameters, including the production PLB allocations that define the distributions of the noisy measurements observed in the DAS DHC implementations. Section \ref{sec:metrics} provides summary metrics for the production 2020 DHC file release. Section \ref{sec:nmfs} concludes with a description of upcoming experimental data products to support statistical inferences on CEF-based tabulations.

\subsection{A Brief Overview of DAS} \label{sec:dasOutline}

Throughout this paper we make frequent references to \textcite{abowd20222020}, but, to make this paper more self-contained, this section outlines the basic steps taken during the DHC and redistricting data DAS executions. First, after reading the microdata contained in the CEF, these implementations use these confidential microdata to create a flattened fully saturated contingency table, \textit{i.e.}, a histogram, for each census block. In other words, each element (or cell) in the histogram of a given census block corresponds to a possible value of the records, as defined by the product schema, and the value of the cell provides the number of records with this value. Since microdata and histogram data formats are referenced throughout this paper, the next example describes these formats and how one can map between these two encodings of the data.

\begin{table}[h!]
    \caption{An example dataset in microdata format}  \label{tab:eg_microdata}
    \begin{tabular}{lll}
    STATE\_GEOCODE & HISPANIC & SEX \\ \hline
    24 & 0 & 0  \\
    24 & 1 & 0  \\
    24 & 1 & 1  \\
    55 & 0 & 1  \\
    55 & 0 & 1 
    \end{tabular}
\end{table}

\begin{example} \label{eg:dataFormats}
    In this example we describe how to map the dataset given in Table \ref{tab:eg_microdata} from microdata format to a histogram. We use \textit{microdata} to mean the data format in which each row provides an observation and each column provides the attribute values of the observations. We refer to the list of attributes and the universe of possible levels (or values) for each attribute as the \textit{schema} of the data. For example, the dataset in Table \ref{tab:eg_microdata} has the following schema.
    \begin{enumerate}
        \item STATE\_GEOCODE. 51 levels: Provides the FIPS state geographic level identification code. Note that the District of Columbia is included as a state equivalent in this geographic level.
        \item HISPANIC. 2 levels: Indicates if the respondent is Hispanic/Latino, or not Hispanic/Latino.
        \item SEX. 2 levels: Indicates if the respondent is Male or Female.
    \end{enumerate}
    In contrast, a histogram provides the count of records with each possible combination of attribute levels. Since the DAS implementation uses histograms of individual geographic units, next we describe how to convert this dataset to the histogram encoding for the state with a STATE\_GEOCODE of 24. This can be done by first filtering the microdata dataset given in Table \ref{tab:eg_microdata} to the three rows with STATE\_GEOCODE$ = 24.$ Second, we sequentially consider each possible combination of the levels of the remaining attributes in lexicographic order, and collect the count of records with the corresponding combination of levels into a vector, which provides the final histogram for this geographic unit. For example, the first element of this lexicographic ordering on the levels of the HISPANIC and SEX attributes are cells with HISPANIC=0 and SEX=0. Since there is one record with these levels, the first such record count is 1. After continuing this process for the remaining level combinations, we can see that the histogram for this geographic unit is given by $\boldsymbol{x} = (1, 0, 1, 1)^\top.$
    
    Histograms are a convenient format to use within the DAS implementation because they allow for many operations to be specified with linear algebra. For example, if $A\otimes B$ denotes the Kronecker product of the matrices $A$ and $B,$ we can use the histogram $\boldsymbol{x}$ to compute the SEX marginal for this geographic unit using 
    \[\left(\begin{bmatrix}
    1 & 1
    \end{bmatrix} \otimes \begin{bmatrix}
    1 & 0 \\
    0 & 1 
    \end{bmatrix} \right)\boldsymbol{x} = \begin{bmatrix}
    1 & 0 & 1 & 0\\
    0 & 1 & 0 & 1
    \end{bmatrix} \begin{bmatrix}
    1\\
    0\\
    1\\
    1
    \end{bmatrix} = \begin{bmatrix}
    2\\
    1
    \end{bmatrix}. \]
    This feature is particularly important when specifying linear constraints in optimization models, in which case the histogram vectors are variables that are being optimized over rather than fixed constants, since all constraints used within DAS other than some integrality constraints can be specified as linear constraints.
\end{example}

After these histograms have been created in each block geographic unit, a process called spine optimization is carried out. There are many details related to this process that we do not cover here but that are described by \textcite{cumings2022geographic}. For the purposes of this paper, it suffices to know that this process defines a hierarchical set of geographic units, which is known as a geographic spine, or simply a spine. For example, the geographic levels for the DHC DAS implementation, from least to most granular are:  US, state, county, prim, tract subset group (TSG), tract subset (TS), optimized block group (OBG), and census block geographic levels. For more detail on the definitions of these geographic levels, see \textcite{cumings2022geographic}.\footnote{State includes state equivalents, namely the District of Columbia. County includes county equivalents. Throughout the paper the geographic unit ``block'' always means ``census block.'' The geographic levels used in DHC Puerto Rico executions are the same as the geographic levels listed here, with the exception that a US geographic level is not included. Prim geographic units are defined as the most granular geography unit used internally by the Population Estimates and Projections (PEP) Area of the Census Bureau. This geographic level includes entities defined as the intersections of counties and other geographic units, such as functioning places and minor civil divisions.} The DHC DAS executions used different spines than the ones used for the redistricting data file DAS executions for two reasons. First, the maximum number of child geographic units of any parent geographic unit is lower for the DHC optimized spine; this is required to ensure that the optimization solves used within DAS executions, which are described below in more detail, are computationally feasible. Second, the differences in use cases between the DHC and redistricting data file releases led to updates to the off-spine geographic entities that the spine optimization routines target accuracy in. For example, one use case for the DHC tabulations is forecasting school enrollment figures, so school districts were added to this set of off-spine entities that are targeted for accuracy in the DHC DAS implementation. Even though the geographic spine used internally within the DAS implementation does not correspond to the Census Bureau Geography Division’s standard tabulation spine, the published 2020 DHC tabulations are still counts for geographic entities on the standard tabulation spine. 

After the optimized spine is created, the histogram for each geographic unit (e.g., Fairfax County, Virginia) in the optimized spine is created by summing up the histograms of the blocks it contains. The DAS noisy measurements module,  a formally private mechanism that satisfies $\rho$-zero-concentrated differential privacy ($\rho$-zCDP) \parencite{bun2016concentrated}, then forms various marginals from these resulting histograms and adds zero-mean discrete Gaussian noise \parencite{canonne2020discrete} to these marginals. These noisy marginals are called \emph{noisy measurements}.

The DAS postprocessing module uses the noisy measurements to create the MDF. Starting from the least granular geographic unit, \textit{i.e.}, the root geographic units, which are either the US or Puerto Rico, it fits a nonnegative and integer histogram to the noisy measurements computed at that geographic level by solving a series of optimization problems, which are described in more detail by \textcite{abowd20222020}. The DAS postprocessing module then fits nonnegative integer histograms to the child geographic units using their noisy measurements, subject to the constraint that their fitted histograms sum to the fitted histogram  of the parent geographic unit. This process is then repeated to define the histogram estimates for each of the remaining geographic levels, from the least to the most granular geographic units. Defining these estimates as solutions to optimization problems also provides a straightforward method to ensure these estimators satisfy a set of constraints, as described by \textcite{abowd20222020}. The specific optimization problems that are used to define the histogram estimate of each geographic unit are carried out in a series of \textit{optimization passes}, each consisting of one nonnegative least squares (NNLS) solve to estimate a vector of non-negative count estimates and one mixed integer programming (MIP) solve to convert these estimates to integers. These optimization passes are repeated to estimate and fix nonnegative and integer counts of progressively more granular tabulations, and the final optimization pass estimates and fixes the detailed histogram counts of the geographic unit. The final step within the DAS implementation is to convert the block geographic level fitted histograms to microdata, which is the output MDF of the DAS implementation. 

After the production DAS executions were complete, the 2020 Census tabulation tables were created using the output MDFs. Part of our planning process for the 2020 Census statistical data product publications involved updates to the tables that were released as part of the 2010 Census statistical data products. This included adding, removing, and altering the granularity of the published tables, as well as moving some tables to the more specialized \textit{Group II statistical data products}, which are the 2020 Census Detailed DHC-A, Detailed DHC-B, and the S-DHC statistical data products. In aggregate, these updates resulted in a reduction in the total number of tables published as part of all statistical data products released for the 2020 Census. The updates that were most relevant to our adoption of the DHC DAS implementation for the 2020 Census were the removal of the tables included in the 2020 Census Group II statistical data products from the DHC publications because supporting tabulations that included the detailed race and ethnicity categories considered in these Group II products (\textit{e.g.}, German, Chinese, Mexican, \textit{etc.}, as opposed to the less granular race and ethnicity categories used in DHCH/DHCP schemas like White, Asian, Hispanic/Latino, \textit{etc.}) would have significantly increased the dimensions of the optimization models in the DHC DAS implementation. \textcite{tableCrosswalk} provides a description of the tables used for the 2020 Census publications, along with a crosswalk between the 2010 Census and the 2020 Census published tables.

\section{DHC Schemas and Constraints}
\label{sec:schemas_and_constrs}

In the following subsections we begin by describing the schemas and the constraints of both DHCH and DHCP. Both of these schemas are more granular than their respective redistricting data counterparts. We also ensure that the DHCP MDF is consistent with the MDF of the persons redistricting DAS output and that the DHCH MDF is consistent with the MDF of the units redistricting DAS output. In other words, it is possible to recreate the redistricting persons universe MDF using only the DHCP MDF and to recreate the redistricting units universe MDF using only the DHCH MDF. Constraints that ensure this consistency between these statistical data products are imposed for two reasons. First, it avoids the possibility of confusion caused by conflicts between these two data products. Second, these constraints allow us to ensure accuracy of the less-granular queries that are fixed by the redistricting data file DAS evaluations without expending additional PLB. Constraining to the redistricting data file production DAS execution outputs also ensures that the state total population invariant constraints that were imposed in these DAS executions hold for tabulations based on the DAS DHC implementation.

The CEF is created using edit and imputation methods to transform the unedited data collected during the decennial Census to data that satisfies certain conditions, called \textit{edit constraints}, and the remaining constraints imposed in DHC DAS executions described in this section are taken directly from these edit constraints. For example, one such edit constraint ensures  that the age of at least one person in each household is at least 15.

\subsection{DHCH Schema} \label{sec:dhch_schema}

As described in Section \ref{sec:intro}, the published DHCH tabulations provide counts for two distinct categories of entities: households and of housing units/GQs. Note that the structure in which a household resides is called a housing unit, and people residing in GQs are not viewed as household members. The majority of the published DHCH tabulations are in the first of these two categories, \textit{i.e.}, they provide counts of households. Since each household has exactly one householder, these counts can alternatively be interpreted as tabulations of householders in certain types of households. While these two categories of tabulations are distinct from one another, they are also related in a few ways. For example, the number of housing units that are owned by their residents with a mortgage in a geographic unit must be equal to the number of households that own their home with a mortgage. These relationships between these two categories of counts provide a way for the noisy measurements of counts of housing units to be used to enhance the precision of count estimates for households. In order to do this, the DHCH DAS implementation estimates counts of these two categories within a single DAS execution by estimating and fixing a secondary \textit{units histogram}, in addition to the main \textit{household histogram}. This section describes the schemas of these two histograms, starting with the main household histogram. To do so, we first define one attribute of this main histogram, called HHTYPE, \textit{i.e.}, \textit{household type}. Each level of HHTYPE is defined as a combination of the following attributes that is actually feasible in practice, which we describe below in more detail. 

\begin{enumerate}
    \item SIZE. 7 levels: Indicates if the household includes 1, 2, $\dots,$ or 7+ members.
    \item SPOUSE. 5 levels: Indicates if the householder is not living with a spouse or partner, is married to and living with someone of the same sex, is married to and living with someone of the opposite sex, is cohabiting with a same sex partner, or is cohabiting with an opposite sex partner.
    \item RELATIVE. 2 levels: Indicates if a relative of the householder is present in the household or not.
    \item OWN\_CHILD. 4 levels: Indicates if the householder does not have an own child (either adopted or biological) less than 18 years old present in the household, has an own child less than 6 but does not have an own child who is 6-17, does not have an own child less than 6 but does have an own child who is 6-17, or if the householder has an own child less than 6 years old and also has an own child who is 6-17 years old. 
    \item CHILD. 4 levels: Indicates if the household does not include a child, includes a child less than 6 but does not include a child who is 6-17, does not include a child less than 6 but does include a child who is 6-17, or if the household includes a child less than 6 years old and also includes a child who is 6-17 years old.
    \item MULTIGEN. 2 levels: Indicates if the household is multigenerational or not. A household is considered multigenerational if at least three parent/child generations are present. Households that include either the child and the grandchild of a householder, or include a child and a parent of the householder, are considered multigenerational.
\end{enumerate}

The DHCH main histogram schema encodes this information in the single attribute HHTYPE because many of the components that make up HHTYPE are related to one another, and encoding the combination of these variables using a single attribute provides a way to exclude combinations of these variables that are impossible in practice. For example, if the SIZE component of HHTYPE indicates the household is composed of only the householder, then SPOUSE must indicate the householder does not live with a spouse or partner, RELATIVE must indicate the household does not contain a relative of the householder, OWN\_CHILD must indicate the householder does not have an own child, and MULTIGEN must indicate the household is not a multigenerational household. In other words, most elements of the cross product of each of these individual components do not actually correspond to a possible level of HHTYPE. Rather than include all 2,240 elements in this na\"ive cross product of these components of HHTYPE, we decrease the dimension of the resulting DHCH histogram by only including the 522 valid cross product elements. 

The DHCH householder histogram schema is defined by the following attributes and levels.

\begin{enumerate}
    \item GEOCODE. 5,892,698 levels: Provides the block geographic identification code.\footnote{This is the number of blocks in 2020 with at least one housing unit or at least one occupied GQ.}  
    \item SEX. 2 levels: Indicates if the householder is Male or Female.
    \item HHAGE. 9 levels: Indicates if the age of the householder is between 15 to 24 years, 25 to 34, 35 to 44, 45 to 54, 55 to 59, 60 to 64, 65 to 74, 75 to 84, or over 85 years.
    \item HISPANIC. 2 levels: Indicates if householder is Hispanic/Latino, or not Hispanic/Latino.
    \item RACE. 7 levels: Indicates if householder is Black/African American, American Indian/ Alaska Native, Asian, Native Hawaiian/Pacific Islander, White, some other race, or two or more of these race categories.
    \item ELDERLY. 4 levels: Indicates if the age of the oldest member of household is less than 60 years, between 60 and 64, between 65 and 74, or over 75 years.
    \item TENURE. 3 levels: Indicates if household resides in a housing unit that is owned with a mortgage, owned without a mortgage, or rented.
    \item HHTYPE. 522 levels: These levels are described in the previous enumerated list.
\end{enumerate}

The DHCH units histogram has a schema with only one attribute other than GEOCODE, which is known as TENVACGQ (35 levels). TENVACGQ provides the tenure or vacancy status for each housing unit and the detailed GQ type category for each GQ. Specifically, the first four levels of TENVACGQ correspond to levels of occupied housing units; these levels indicate if a property is mortgaged, owned, rented, or occupied without payment of rent. The subsequent seven levels correspond to types of vacant housing units; these levels indicate if the housing unit is vacant and for rent, vacant and rented but not occupied, vacant and for sale, vacant and sold but not occupied, vacant because the housing unit is for seasonal, recreational, or occasional use, vacant because the property is for migrant workers, or vacant for other reasons. The subsequent 24 levels correspond to the detailed GQ type categories used in DHCH. The units histogram is estimated at the same time as the householder histogram during the DHCH DAS implementation because a constraint links both of these histograms to one another. This constraint, and all other constraints in the 2020 DHCH production implementation settings, are provided in the list below.

\begin{enumerate}[label=(H\arabic*), ref=C\arabic*]
    \item The DHCH MDF is constrained to be consistent with the units redistricting data MDF.\label{enum:dhch1}
    \item The count of the total number of housing units and number of occupied GQs is held invariant in each block.\footnote{Note that the counts of occupied GQs are held invariant in each block, but, for housing units, only the counts of housing units are held invariant in each block. In other words, both the number of vacant and occupied housing units in each block are protected, but their sum is held invariant.} \label{enum:dhch2}
    \item  Structural zero constraints are included to ensure consistency of attributes with the SIZE component of HHTYPE. For example, households of size one cannot have a householder with age less than 60 and have a household member with age greater than 60.\label{enum:dhch3}
    \item The values of ELDERLY, HHAGE, and HHTYPE are constrained to ensure that the difference between the age of each householder and the implied age of their spouse/partner is no more than 50 years. For example, if the HHAGE attribute indicates the householder is between 15-24 years old and HHTYPE implies that the household contains only the householder and their spouse/partner, then ELDERLY cannot imply that a member of the household is older than 75 years of age.\label{enum:dhch4}
    \item The value of ELDERLY is constrained to be at least as high as HHAGE for each record. \label{enum:dhch5}
    \item Age differences between parents and children are constrained to be less than the upper bound on these differences that is used in the CEF edit constraints. For example, these constraints rule out households with a householder who is over 75 years old and that includes a child of the householder who is less than 6.\footnote{The CEF edit constraints bound the maximum age difference between a mother and a child to 50 years and the maximum age difference between a father and a child to 69 years.} \label{enum:dhch6}
    \item The DHCH householder and unit histograms are constrained to be consistent with one another; the count of householders who own their home with a mortgage, own without a mortgage, and do not own their home in the householder histogram are each constrained to be equal to the count of homes that are owned with a mortgage, owned without a mortgage, and that are not owned by the current occupants in the units histogram, respectively. \label{enum:dhch7}
\end{enumerate}

\subsection{DHCP Schema} \label{sec:dhcp_schema}

Unlike the DHCH DAS executions, there is only one category of counts that are estimated in DHCP DAS executions, which are counts of people, including both people in housing units and GQs, so DHCP DAS executions only use one histogram. Specifically, the DHCP schema is defined by the following attributes and levels.

\begin{enumerate}
    \item GEOCODE. 5,892,698 levels: Provides the block geographic identification code.  
    \item RELGQ. 42 levels: For respondents in households, the relationship to householder is encoded by the first 18 levels, and, for the remaining respondents, the detailed GQ type is encoded by the final 24 levels.
    \item SEX. 2 levels: Indicates if the respondent is Male or Female.
    \item AGE. 116 levels: Contains one level for each single year of age between 0 and 115. 
    \item HISPANIC. 2 levels: Indicates if the respondent is Hispanic/Latino, or not Hispanic/Latino.
    \item CENRACE. 63 levels: Includes every combination of Black/African American, American Indian/Native Alaskan, Asian, Native Hawaiian/Pacific Islander, White, and some other race, except ``none of the above.''
\end{enumerate}

The invariant constraints for the 2020 DHCP DAS implementation are described in the following list. 

\begin{enumerate}[label=(P\arabic*), ref=P\arabic*]
    \item The DHCP MDF is constrained to be consistent with the persons redistricting data, \textit{e.g.}, recomputing any tabulation that is in the redistricting data file using the DHCP MDF results in the same values as those in the redistricting data file. \label{enum:dhcp1}
    \item The count of occupied GQs for each detailed GQ type in each block is held invariant. In other words, for a given block and a given detailed GQ type, we require at least the same number of records in this detailed GQ type and block as the number of occupied GQs of this type in the block. In addition, the number of records in each GQ is constrained to be at most 99,999, and the number of records in each housing unit is also constrained to be at most 99,999. \label{enum:dhcp2}
    \item Depending on the value of RELGQ of each record, the AGE attribute of the record is constrained to satisfy the lower and upper bounds given in Table \ref{table:dhcp_age_feas}, which are also edit constraints that are used when creating the CEF.  \label{enum:dhcp3}
\end{enumerate}

\begin{table}[h!]
    \centering
    \begin{tabular}{ p{12cm}|p{2cm}}
        Relationship to Householder or GQ Type & Age Range \\
        \hline 
        Householder  & 15-115 \\
        Spouse or unmarried partner of householder   & 15-115  \\
        Adopted or Biological child  & 0-89  \\
        Stepson/stepdaughter  & 0-89 \\ 
        Brother or sister  & 0-115  \\
        Parent  & 30-115  \\
        Grandchild  & 0-74  \\
        Parent-in-law  & 30-115 \\ 
        Son-in-law/daughter-in-law  & 15-89  \\
        Other relative  & 0-115  \\
        Housemate/roommate  & 0-115 \\ 
        Foster child  & 0-20  \\
        Other non-relative  & 0-115 \\ 
        Non-military correctional facilities (GQ Types 101-105)  & 15-115  \\
        Group home/correctional/treatment facilities for juveniles (GQ Types 201-203)  & 0-25  \\
        Nursing facilities (GQ Type 301)  & 20-115  \\
        Hospitals/hospices (GQ Types 401-403)  & 0-115  \\
        Residential schools for people with disabilities (GQ Type 405)  & 3-30  \\
        College/university student housing (GQ Type 501)  & 16-65  \\
        Military Housing (GQ Types 106, 404, 601, 602)  & 17-65  \\
        Adult group homes/treatment centers (GQ Types 801, 802)  & 16-115  \\
        Maritime/merchant vessels (GQ Types 900)  & 16-75  \\
        Other non-institutional (GQ Types 701-706, 901-904)  & 0-115 \\ 
    \end{tabular}
    \caption{This table provides the feasible age ranges by relationship to householder and GQ type.}
    \label{table:dhcp_age_feas}
\end{table}

\subsection{Types of Constraints Within the DAS}

For both the DHCH and DHCP DAS implementations, it is worthwhile to separate constraints into two categories. First, \textit{edit and structural zero constraints} are imposed on the output of the DAS to ensure that the constraints imposed during the creation of the CEF are also satisfied in the output MDF. These constraints can be viewed as the requirements provided by SMEs for the face validity of the microdata. These constraints are also data-independent and publicly available. As a result, and also since our privacy guarantees are specified with respect to the records in the CEF, these constraints do not limit the types of inferences on the CEF records that we protect against. Some examples of constraints in this category include that each householder record in the DHCP MDF is constrained to be at least 15 years of age, and, in the DHCH MDF, that each household of size one with a householder with age less than 60 is constrained to have ELDERLY indicate that no household member is over 60 years of age. Other examples of these constraints include constraints (\ref{enum:dhch3}-\ref{enum:dhch7}) and (\ref{enum:dhcp3}) in the list of DHCH and DHCP constraints above, respectively.

Second, our production DHC implementations also impose several constraints, which we call \textit{invariant constraints}, that are dependent on the input data files of the DAS DHC implementations rather than on publicly-available data. Specifically, the invariant constraints imposed on the production DHC implementations are (A) the total populations of each state, Washington D.C., and Puerto Rico, (B) the tabulation block locations of each \textit{occupied} GQ by detailed GQ type, \textit{e.g.}, the location of each nursing facility, and (C) the tabulation block locations of each housing unit. The first set of constraints is imposed in the persons universe of the production 2020 Redistricting Data File DAS implementations. This is carried forward into   the production DHCP implementation through our constraints that ensure the DHC outputs are consistent with the redistricting data file. While we  protect inferences on whether a given a housing unit is occupied or vacant, we do not provide this same protection to GQs because unoccupied GQs are out of scope for the 2020 Census and the DAS has no access to that information. More detail on the rationale for the inclusion of these invariant constraints is provided by \textcite{abowd20222020}.

It is possible to use these set of invariant constraints to formulate inferences on individuals that are not protected by the DAS. For example, if a privacy attacker knows an individual resides in a given block that does not have an occupied nursing facility in the DHCH units table, it is possible for the privacy attacker to infer that the individual is not in a nursing facility.

\subsubsection{Unused Constraints in DHC}

Note that the constraints imposed within the DHC DAS implementation do not avoid all inconsistent counts. For example, since the persons and units universe are processed separately from one another, there are some blocks with occupied housing units in the DHCH MDF but without any population in the DHCP MDF. Inconsistent counts such as these are most common in sparsely populated areas. Ensuring consistency between the persons and units universes using an approach that is similar to the optimization passes that are currently used in the DAS implementation might not be computationally feasible because this would significantly increase the size of the optimization problems solved within DAS.\footnote{The next section also describes the sufficient condition that we use to ensure our MIP solves are computationally tractable. This sufficient condition would also fail to hold if we added the constraints required for consistency between the counts in these two universes to combined optimization solves that estimate and fix the DHCH and DHCP histograms simultaneously.} Also, it is worth noting that inconsistent counts are an accepted attribute of census tabulations in other countries, such as in the UK Census \parencite{doveApplying,ONS:2023}.  

\subsubsection{How Constraints are Imposed}

As described previously, the primary way that fitted histograms are computed to satisfy the DHCH and DHCP constraints is through a series of least squares and mixed integer programming optimization solves. These optimization problems are described in detail by \textcite{abowd20222020}. As is also described in \textcite{abowd20222020}, we require  the constraint matrix for the MIP optimization problems to satisfy a condition  known as total unimodularity. This restriction ensures these MIP solves are computationally tractable; for more detail on total unimodularity and the sufficient condition we use to ensure it, see \textcite{abowd20222020, asano2003matrix}.

With the exception of the DHCP constraints described in (\ref{enum:dhcp2}), \textit{i.e.}, the upper and lower bound constraints on populations in each GQ, all other DHCH and DHCP constraints are imposed by including them directly in the optimization solves. A different approach is required for DHCP constraints described in (\ref{enum:dhcp2}) because the sufficient condition we use to ensure total unimodularity would not hold if these constraints were added directly to the MIP optimization problems. For this reason, we created a post-processing method called the \emph{DHCP optimization heuristic} that improves the feasibility for the constraints in (\ref{enum:dhcp2}) of the histogram estimate that is output from the optimization passes described in Section \ref{sec:dasOutline}. More specifically, this method ``improves feasibility" of the upper and lower bound constraints in (\ref{enum:dhcp2}) on the detailed GQ population counts by moving each detailed GQ population count closer to the interval of feasible GQ population counts, but it does not ensure that all of these constraints are actually satisfied in all cases, as we describe below in more detail. This method works by iteratively moving records between pairs of histogram cells for each individual child geographic unit, with these pairs defined so that these moves maintain feasibility of the DHCP constraints described in (\ref{enum:dhcp1}) and (\ref{enum:dhcp3}).

Before describing this optimization heuristic, some additional background will be helpful. Note that the redistricting data file persons MDF was constrained to satisfy similar bounds on the population within each major GQ type.\footnote{The major GQ type code is given by the first digit of the three detailed GQ type code for each detailed GQ type code less than 700, and is equal to 7 for all remaining detailed GQ types. For example, the major GQ type code for detailed GQ type 301, \textit{i.e.}, nursing facilities, is 3, and the major GQ type code for detailed GQ type 900, \textit{i.e.}, maritime/merchant vessels, is 7.} To continue to satisfy the constraints described in (\ref{enum:dhcp1}), \textit{i.e.}, the constraints that ensure consistency with the redistricting data file persons tabulations, while moving the population with a given detailed GQ type toward a feasible interval, we only need to consider moving records between pairs of histogram cells with the same major GQ type but different detailed GQ types. Throughout the discussion below we will consider an example of a geographic unit with one maritime/merchant vessels (GQ type 900, which is in major GQ type 7) and no other GQs. We will also suppose that the histogram estimate output from the optimization passes places only one estimated resident in a GQ type, that this person is placed in an emergency and transitional shelter (GQ type 701, which is also in major GQ type 7), and that this respondent is estimated to be 10 years old. This is an interesting example for a few reasons. First, in this case, the constraints described in (\ref{enum:dhcp2}) require that between 1 and 99,999 records in this geographic unit reside in GQ type 900 and that no one is residing of GQ type 701, so the histogram estimate output from the optimization passes does not satisfy two constraints in (\ref{enum:dhcp2}). Second, simply switching the GQ type of the sole record estimated to be in major GQ type 7 from 701 to 900 would result in one constraint on AGE in (\ref{enum:dhcp3}) failing to hold, since each record in GQ type 900 is required to be at least 16 years old.

The DHCP optimization heuristic is run separately on the histogram estimate of each individual child geographic unit, and the method does not ensure that parent-child consistency is maintained. In the example described above, the optimization heuristic may move the sole estimated GQ resident from GQ type 701 to GQ type 900 within this geographic unit. However, since the method is run separately on the child geographic units of each parent without requiring that the changes to the histograms of each child geographic unit cancel with one another, it is likely that the method changes the count of records in these two GQs in total over all of the child geographic units, so these two counts are not likely to be consistent with the previous estimate of the parent geographic unit. Since the TopDown Algorithm only uses the previously estimated geographic level when estimating the next most granular geographic level (rather than using histogram estimates from all previously estimated geographic levels), the typically-small number of record changes carried out by the optimization heuristic each time it is called has the potential to cause larger changes in tabulations of marginals in less-granular geographic units, such as the US geographic unit, after it is applied to the histograms of each of the more granular geographic levels. Below we will describe the issues caused by not ensuring parent-child consistency, after describing the method in more detail.

This heuristic works by iteratively considering each detailed GQ type that is included as a level of the attribute RELGQ. When the histogram counts for this level of RELGQ are satisfied, \textit{i.e.}, the estimate of individuals in this detailed GQ type are between the required lower and upper bounds, the child histogram is not altered and the next detailed GQ type is considered. Otherwise, the optimization heuristic defines a set of pairs of histogram cells, with one element of each pair called the donor cell and the other the recipient cell, and with each pair defined so that moving a record from the donor cell to the recipient cell would move the population estimate for the detailed GQ type being considered toward the interval of feasible counts and also such that certain other conditions hold, which we describe below in more detail. If there exists at least one such pair of donor and recipient cells that provide a way to move a record without violating any of the remaining DHCP constraints, a pair is chosen uniformly at random from these possible pairs and then a record from the chosen donor histogram cell is moved to the chosen recipient histogram cell. This process is repeated until the histogram estimate satisfies the detailed GQ bound constraint or there does not exist a possible record to move without violating one or more other DHCP constraints, at which point the next detailed GQ bound constraint is considered. 

The additional conditions that are used to define the pairs of donor and recipient histogram cells are set to balance a tradeoff between feasibility and accuracy of the tabulations. For example, we could ensure all of the counts satisfy all of the DHCP constraints described in Section \ref{sec:dhcp_schema} by setting this condition so that all donor and recipient cells have identical values for all attributes other than RELGQ and AGE. In particular, this goal would require allowing for changes in respondents' AGE attribute (in addition to changes in RELGQ) to restore feasibility of the constraints described in (\ref{enum:dhcp3}), in the event that the set of feasible ages for a record change due to change in its RELGQ attribute. Returning to our running example, as described above, if we were to change the GQ type of the sole record estimated to be in a GQ from 701 to 900, then maintaining feasibility of the constraints described in (\ref{enum:dhcp3}) would require that we also change the AGE attribute of this record to at least 16. After testing this approach of ensuring feasibility of all DHCP constraints, we found that this resulted in too large of a perturbation in the AGE marginals. Since the optimization heuristic does not ensure parent-child consistency, the aggregate impact of the optimization heuristic on the AGE marginal estimate for the US as a whole can be derived by summing over the changes to the AGE marginals made by the optimization heuristic in all geographic units. For this reason, requiring feasibility of all constraints in each geographic unit was especially problematic for the accuracy of the AGE marginal for the US as a whole. To ameliorate this issue, the DHCP production settings require that all donor and recipient record pairs in the optimized block group and block geographic levels have identical levels for every attribute other than RELGQ. In other words, this optimization heuristic does not consider changes to the AGE marginal in these two most granular geographic levels. In terms of our running example, if this child geographic unit is in either of these two most granular geographic levels, the optimization heuristic would not make any changes to the estimated output from the optimization passes, since doing so would require a change in the AGE attribute of this record. 

In all geographic levels higher than the optimized block group and block geographic levels on the spine, we first attempted to define the donor and recipient cell pairs in the same way that is done for these two most granular geographic levels, and only resort to changing the AGE levels of records when this approach is not capable of restoring feasibility of all the constraints described in (\ref{enum:dhcp2}). So, if the child geographic unit in our running example is in these less-granular geographic levels, the optimization heuristic would return a histogram with the sole record estimated to be in a GQ having its GQ type changed from 701 to 900 and its AGE attribute changed from 10 to either 16 or 17 (both with a probability of $0.5).$ Note that the updated AGE attribute of either 16 or 17 results in both the AGE of the residents of GQ type 900 continuing to be feasible, \textit{i.e.}, feasibility of the constraints in (\ref{enum:dhcp3}), and, since VOTING\_AGE is one attribute included in the persons redistricting data file MDF, the histogram continuing to be consistent with the previously estimated persons redistricting data file MDF, \textit{i.e.}, feasibility of the constraints in (\ref{enum:dhcp1}). In practice, we found that these settings resulted in both the constraints described in (\ref{enum:dhcp2}) typically being feasible and the AGE and RELGQ marginals of the less-granular geographic units, such as the US as a whole, being sufficiently accurate. 

In terms of runtime, this approach is not a major bottleneck in DAS DHCP executions because the number of records moved between cells is typically not high for our production settings. This is a result of several factors, including the fact that the histogram estimates returned by the optimization passes for most child geographic units are guaranteed to satisfy the constraints in (\ref{enum:dhcp2}) for a few reasons. Specifically, most geographic units do not have any GQs, especially those at granular geographic levels, and the population in each detailed GQ type for these geographic units for the histogram estimates returned from the optimization passes is constrained to their correct value of zero for these geographic units by the constraints in (\ref{enum:dhcp1}), since major GQ type was estimated in the persons redistricting data file MDF. Also, even when this is not the case, our 2020 DAS DHCP production settings typically only result in modest errors for each detailed GQ type population, and, since the CEF-based GQ type populations satisfy the constraints in (\ref{enum:dhcp2}) for each geographic unit, the accuracy of the histogram estimate returned by the optimization passes has the effect of limiting the number of records that are moved by the DHCP optimization heuristic. 

\section{Computational Requirements for the DHC DAS} \label{sec:implementation_details}

\textcite{abowd20222020} provide detail on the computational environment used by the redistricting data file DAS implementation, including a list of software dependencies, AWS cluster configurations, and Spark settings. In this section, we describe the computational environment used for our DHC DAS implementations, with a particular focus on the aspects that we changed after publication of \textcite{abowd20222020}. 

The factors that did not change include the software dependencies, \textit{i.e.}, our production code base is written in Python version 3.7, we use Gurobi 9.1, and the system ran on  Amazon Web Services' (AWS) using Elastic Map Reduce (EMR) version 6.2 and Apache Spark version 3.0.1. The DAS is run on EMR clusters built from AWS nodes with 96 cores and 768 GiB of RAM. The DHC executions were carried out on clusters with 20 (core and task) worker nodes and one EMR master node. 

Several changes were also made to avoid computational issues due to the DHC schemas being much larger than their redistricting data counterparts. First, due to the memory requirements of the large optimization solves used in DHC executions, we increased executor memory and overhead memory to 423 GiB and 330 GiB, respectively. Our typical DHC executions used 20 executors, 40 cores per executor, and 40 GiB of driver memory. We also made several changes to our cluster configuration settings, including increasing the heap sizes of the yarn node manager and resource manager to 8 GiB and 4 GiB, respectively, increasing the EBS storage to 16 TiB, and switching to a G1 garbage collection method.

\section{DHC Accuracy Experiments and PLB Allocations} \label{sec:accuracy_experiments}

In preparation for the 2020 DHC production DAS implementations, the DAS team performed a series of accuracy experiments using 2010 Census inputs in order to support recommendations for the Census Bureau's final production settings of the 2020 DHC production implementations, including the PLB allocations. The DHC accuracy experiments were similar to those used to support the final production settings for the 2020 Redistricting Data (P.L. 94-171) Summary File production DAS implementations \parencite{abowd20222020}. Specifically, these accuracy experiments used an iterative process consisting of DAS configuration updates based on input from SMEs on the accuracy of tabulations from prior experiment iterations. The Census Bureau internal SMEs included those from the Population Division (POP), and Social, Economic, and Housing Statistics Division (SEHSD). 

External SMEs also provided feedback based on the output of two DHC demonstration data products (DDPs) based on 2010 Census input data released on March 16, 2022 and August 25, 2022, respectively.\footnote{Despite the name ``demonstration data products," these releases were not meant to demonstrate the final production settings we intended to use. These were instead intended to demonstrate performance of the DHC DAS implementation for the settings we were currently testing at the time in our DHC accuracy experiments.} This includes feedback provided through a National Academies of Science, Engineering, and Medicine's Committee on National Statistics (CNSTAT) meeting in June of 2022 \parencite{cnstat}, academic conferences, and external stakeholders independently contacting the Census Bureau regarding the accuracy of these 2010 DDPs for specific use cases. In addition to these two DDPs, the DAS output based on 2010 Census input data and using the 2020 DHC production implementation settings was also released on April 3, 2023. All of these releases are currently available at \parencite{DHCddps}. 

As was also described by \textcite{abowd20222020} in the context of tuning the DAS for the 2020 production redistricting data file DAS executions, a significant portion of the time that went into tuning DAS during the DHC accuracy experiments focused on collecting accuracy targets  for DHC use cases provided by internal and external SMEs and then attempting to satisfy these targets with minimal PLB. In other words, one of the main focuses during these accuracy experiments was solving the dual problem to the more standard problem in the formal privacy literature of maximizing data utility such that the PLB is below a prespecified fixed threshold. This is in part because the majority of the feedback we received over the course of these accuracy experiments was related to concerns with data quality rather than confidentiality. A wide variety of accuracy metrics were used over the course of the DHC accuracy experiments, for a variety of queries and sets of geographic entities, including measures of central tendency of the absolute value of residuals (\textit{e.g.}, mean and median absolute error), quantiles of residuals, and measures of absolute relative residuals. The DAS team also used privacy-semantic statements, primarily those based on bounds on the statistical power of hypothesis tests on the attributes of CEF records, to communicate privacy concerns \parencite{kifer2022bayesian}. The DAS team also provided DSEP with results from a reconstruction/reidentification attack using the DAS output from our test runs, which is described in more detail by \textcite{abowd20232010}. Ultimately, choosing the production DAS settings that balance the tradeoffs between accuracy, privacy, and the quantity of statistics is a policy decision that is the responsibility of DSEP. DSEP chose the 2020 DHC DAS production settings based on background information, accuracy metrics, and recommendations from internal stakeholders that were based on the results of the DHC accuracy experiments.

The result of these accuracy experiments was the 2020 DHC production configuration parameters, including the DHCH and DHCP \textit{strategy matrices}, which refer to the marginals that are chosen to become noisy measurements as well as the PLB allocation for each marginal, which in turn defines the scale of the noise added to the confidential CEF-based marginal, for each geographic level. As described in \textcite{abowd20222020} in more detail, we use the $\rho$-zero-concentrated differential privacy ($\rho$-zCDP) accounting framework to specify these PLB values \parencite{bun2016concentrated}. Each strategy matrix described in the next two subsections provides the initial PLB allocations for each geographic unit on the spine, prior to the spine optimization routines' alteration of these allocations for parent and child geographic units for which the parent has only one child; for more detail on the spine optimization routines, see\textcite{cumings2022geographic}. Aside from dictating the scale of the noise for each noisy measurement used within the DAS, the PLB allocations are also useful because they can be used to derive upper bounds on the power of certain hypothesis tests on any given individual record in the CEF; for more detail, see \textcite{kifer2022bayesian}. 

\subsection{DHCH Production PLB Allocations } \label{sec:dhch_plb_settings}

The marginal query groups used in the DHCH strategies are provided in Table \ref{table:dhch_queries}. Many of these marginals are defined by the attributes in the Section \ref{sec:dhch_schema}. The definitions of the remaining marginals, \textit{i.e.}, recodes, are given in the list below.

\begin{enumerate}
    \item TENURE\_2LEVELS. 2 levels: Indicates if an occupied housing unit is owned or rented.
    \item PARTNER\_TYPE\_CHILD\_STAT. 12 levels: This HHTYPE attribute marginal can be viewed as the Cartesian product of two variables. The first variable contains three levels that indicate if a householder has no spouse or partner, is married, or is cohabiting with a partner. The second variable is given by the HHTYPE attribute OWN\_CHILD, as defined in Section \ref{sec:dhch_schema}.
    \item HHTYPE\_RECODE. 183 levels: Defined as the cross product of HHTYPE attributes, MULTIGEN$\times$SPOUSE$\times$SIZE$\times$OWN\_CHILD.
\end{enumerate}

\begin{table}[H] 
\caption{Production DHCH Strategy Queries}
\begin{tabular}{|R{1.45cm} |R{11.1cm} |R{1.45cm}|}
\hline
Query ID &  Marginal Query Group Name & Cells \\
\hline 
1H & MULTIGEN$\times$HISPANIC$\times$TENURE\_2LEVELS & 8\\
2H & HISPANIC$\times$TENURE\_2LEVELS & 4\\
3H & HISPANIC$\times$TENURE\_2LEVELS$\times$RACE & 28\\
4H & PARTNER\_TYPE\_CHILD\_STAT$\times$SEX$\times$TENURE\_2LEVELS & 48\\
5H & SPOUSE$\times$HISPANIC$\times$TENURE\_2LEVELS & 20\\
6H & SEX$\times$HISPANIC$\times$TENURE$\times$RACE$\times$HHTYPE\_RECODE & 15,372\\
7H & SEX$\times$HISPANIC$\times$TENURE$\times$RACE$\times$HHAGE$\times$HHTYPE\_RECODE & 1,383,48\\
8H & TENURE$\times$HHAGE$\times$HHTYPE\_RECODE & 4,941\\
9H & SEX$\times$HISPANIC$\times$TENURE$\times$RACE$\times$ELDERLY$\times$HHAGE$\times$HHTYPE & 1,578,528\\
10H & TENVACGQ & 35\\
\hline
\end{tabular} \label{table:dhch_queries}
\end{table}

Each marginal query group in Table \ref{table:dhch_queries} is assigned a query ID, and the PLB allocations for the production US DHCH implementation for each query ID and each geographic level are given in Table \ref{table:dhch_allocations}. The values in the table below are the actual $\rho$ allocations multiplied by 10,000. The total value of $\rho$ allocated to all marginal query groups for our production DHCH settings was 15401/2000 ($\approx 7.70$). 

\begin{table}[H] 
\caption{Production US DHCH $\rho$ Allocations $\times 10,000$ (\textit{i.e.}, divide by 10,000 to obtain the $\rho$ value)}
\begin{tabular}{|R{1.5cm} |R{1.2cm}R{1.2cm}R{1.2cm}R{1.2cm}R{1.2cm}R{1.2cm}R{1.2cm}R{1.2cm}|}
\hline
Query ID & US & State & County & Prim* & TSG* & TS* & OBG* & Block\\
\hline 
1H & 792 & 714 & 986 & 973 & 973 & 1,150 & 291 & 11\\
2H & 212 & 193 & 266 & 214 & 214 & 310 & 78 & 3\\
3H & 212 & 191 & 1,500 & 1,500 & 1,500 & 1,750 & 442 & 17\\
4H & 792 & 1428 & 1,972 & 1,946 & 1,946 & 2,300 & 581 & 22\\
5H & 792 & 714 & 3,944 & 973 & 973 & 4,601 & 1,162 & 44\\
6H & 257 & 3,216 & 1,380 & 973 & 973 & 1,610 & 407 & 15\\
7H & 257 & 3,216 & 1,380 & 973 & 973 & 1,610 & 407 & 15\\
8H & 212 & 193 & 266 & 214 & 214 & 310 & 78 & 3\\
9H & 1,049 & 3,931 & 1,380 & 973 & 973 & 1,610 & 407 & 15\\
10H & 259 & 3,216 & 1,380 & 973 & 973 & 1,610 & 407 & 15\\
\hline
\multicolumn{9}{|p{14.5cm}|}{\footnotesize{*The prim, tract subset group (TSG), tract subset (TS), and optimized block group (OBG) geographic levels are not in the tabulation geographic spine and are only used internally in the TDA.}}\\
\hline
\end{tabular} \label{table:dhch_allocations}
\end{table}

\subsection{DHCP Production PLB Allocations } \label{sec:dhcp_plb_settings}

As described in the previous section for the DHCH schema, the marginal query groups used in the DHCP strategies are provided in Table \ref{table:dhcp_queries}. Many of these marginals are defined by the attributes in Section \ref{sec:dhcp_schema}. The definitions of the remaining marginals, \textit{i.e.}, recodes, are given in the list below.

\begin{enumerate}
    \item AGE\_18\_64. 3 levels: Indicates which of the following categories the respondent's age is in AGE $\leq 17,$ $18\leq$ AGE $\leq 64,$ or AGE $\geq 65.$
    \item RELGQ\_4\_GROUPS. 4 levels: Indicates if the respondent is either a householder, a non-householder member of a household, a resident of an institutional GQ, or a resident of a non-institutional GQ. 
    \item AGE\_38\_GROUPS. 38 levels: Indicates which of the following categories the respondent's age is in AGE $=0,$ AGE $=1, \dots,$ AGE $=14,$ AGE $\in [15,17],$ AGE $\in [18,19],$ AGE $\in [20,24],$ AGE $\in [25,29],\dots,$ AGE $\in [55, 59],$ AGE $\in[60,61],$ AGE $\in[62,64],$ AGE $\in [65,66],$ AGE $\in [67,69],$ AGE $\in[70,74],$ AGE $\in[75,79],\dots,$ or AGE $\geq 110.$ 
    \item GQ\_CONSTR\_GROUPS. 6 levels: Indicates if the respondent is a resident of a GQ with a lower-bound constraint on AGE given by 0, 3, 15, 16, 17 years of age, as described in Table \ref{table:dhcp_age_feas}, or if the respondent is a member of a household. In other words, GQ\_CONSTR\_GROUPS indicates if the respondent's GQ type is in $\{201,202,203,301,401,402,403,701\}, \{405\}, $ $\{101,102,103,104,105\}, \{501,801,802,900,901,702,704,706,903,904\}, \{106, 404, 601,$ $602\},$ or if the respondent is in a household. 
    \item AGE\_10\_GROUPS. 10 levels: Indicates which of the following categories the respondent's age is in AGE $< 3,$ AGE $=[3, 14],$ AGE $=15,$ AGE $=16,$ AGE $=17,$ AGE $=[18,19],$ AGE $\in [20, 24],$ AGE $\in[25,29],$ AGE $\in [30,34],$ or AGE $\geq 35.$ 
    \item POP\_SEHSD\_TARGETS\_RELSHIP. 19 levels: Indicates if the respondent is in a GQ or, for residents who are members of a household, which of the 18 household RELGQ categories the respondent belongs to. 
    \item AGE\_40\_GROUPS. 40 levels: Indicates which of the following categories the respondent's age is in AGE $=0,$ AGE $=1, \dots,$ AGE $=17,$ AGE $\in [18,19],$ AGE $\in [20,24],$ AGE $\in [25,29],\dots,$ AGE $\in [55, 59],$ AGE $\in[60,61],$ AGE $\in[62,64],$ AGE $\in [65,66],$ AGE $\in [67,69],$ AGE $\in[70,74],$ AGE $\in[75,79], \dots,$ AGE $\in[105,109],$ or AGE $\geq 110.$ 
    \item RELSHIP\_GQ. 25 levels: For respondents in GQs, indicates the respondent's major GQ type, and, for respondents who are members of households, indicates the respondent's household RELGQ category.
\end{enumerate}

\begin{table}[H] 
\caption{Production DHCP Strategy Queries}
\begin{tabular}{|R{1.5cm} |R{11cm} |R{1.5cm}|}
\hline
Query ID &  Marginal Query Group Name & Cells \\
\hline 
1P & AGE\_18\_64$\times$RELGQ\_4\_GROUPS & 12 \\
2P & AGE\_18\_64$\times$SEX & 6\\
3P & AGE\_38\_GROUPS$\times$SEX & 76\\
4P & HISPANIC$\times$SEX & 4\\
5P & SEX$\times$RELGQ\_4\_GROUPS & 8\\
6P & GQ\_CONSTR\_GROUPS$\times$AGE\_10\_GROUPS & 60\\
7P & POP\_SEHSD\_TARGETS\_RELSHIP & 19\\
8P & HISPANIC$\times$SEX$\times$AGE\_40\_GROUPS$\times$RELSHIP\_GQ$\times$CENRACE & 252,000\\
9P & RELGQ$\times$AGE\_40\_GROUPS$\times$HISPANIC$\times$CENRACE$\times$SEX & 423,360\\
10P & RELGQ$\times$SEX$\times$AGE$\times$HISPANIC$\times$CENRACE & 1,227,744\\
\hline
\end{tabular} \label{table:dhcp_queries}
\end{table}

The PLB allocations for the production US DHCP implementation for each query ID in Table \ref{table:dhcp_queries} and each geographic level are given in Table \ref{table:dhcp_allocations}. As is done in the previous subsection, the values in the table below are the actual $\rho$ allocations multiplied by 10,000; in other words, the actual $\rho$ allocations used in the DHCP DAS production settings are given by the values in the table below divided by 10,000. The total value of $\rho$ allocated to all marginal query groups for our production DHCP settings was 24811/5000 ($\approx 4.96$). 

\begin{table}[H] 
\caption{Production US DHCP $\rho$ Allocations $\times 10,000$}
\begin{tabular}{|R{1.5cm} |R{1.2cm}R{1.2cm}R{1.2cm}R{1.2cm}R{1.2cm}R{1.2cm}R{1.2cm}R{1.2cm}|}
\hline
Query ID & US & State & County & Prim* & TSG* & TS* & OBG* & Block\\
\hline 
1P & 73 & 999 & 310 & 478 & 478 & 868 & 430 & 11\\
2P & 73 & 999 & 310 & 478 & 478 & 868 & 430 & 11\\
3P & 73 & 999 & 310 & 478 & 478 & 868 & 430 & 11\\
4P & 73 & 999 & 310 & 478 & 478 & 868 & 430 & 11\\
5P & 73 & 999 & 310 & 478 & 478 & 868 & 430 & 11\\
6P & 292 & 3,996 & 1,240 & 1,912 & 1,912 & 3,472 & 430 & 11\\
7P & 73 & 999 & 1,240 & 478 & 478 & 3,472 & 430 & 11\\
8P & 73 & 999 & 310 & 478 & 478 & 868 & 430 & 11\\
9P & 73 & 999 & 310 & 478 & 478 & 868 & 430 & 11\\
10P & 73 & 999 & 310 & 478 & 478 & 868 & 430 & 11\\
\hline
\multicolumn{9}{|p{14.5cm}|}{\footnotesize{*The prim, tract subset group (TSG), tract subset (TS), and optimized block group (OBG) geographic levels are not in the tabulation geographic spine and are only used internally in the TDA.}}\\
\hline
\end{tabular} \label{table:dhcp_allocations}
\end{table}

\section{Summary Metrics for 2020 DHC Production Implementations} \label{sec:metrics}

This section provides $\rho$-zCDP approximations of mean absolute error (MAE) metrics for several important DHC tabulations. The MAE of a query in a given geographic level is defined by first finding the query error in each geographic unit in the geographic level; afterward, the MAE is defined as the arithmetic mean of the absolute value of these errors. After computing the MAE values reported in this section, a discrete Gaussian mechanism was used to ensure the release satisfies $\rho$-zCDP. The PLB allocated to each of the state geographic level MAEs was set so that the reported approximate MAE was within 0.5 units of the unprotected CEF-based MAE value with probability 0.9. Likewise, the PLB allocations of the remaining summary metrics were set so that the reported approximate MAE was within 0.15 units of the unprotected CEF-based MAE value with probability 0.9.  Many other $\rho$-zCDP approximate summary metrics, in addition to the ones that are provided here, are available in \textcite{DHCmetrics}. The total PLB for all the summary metrics in this release was $\rho=0.7.$

\begin{table}[h!] 
\caption{Production US 2020 DHCH: Presence of Own Child of Householder  $\rho$-zCDP MAE}
\begin{tabular}{|R{7.5cm} |R{1.4cm}R{1.4cm}R{1.4cm}R{1.4cm}|}
\hline
Group & States & Counties & Places & Tracts \\
\hline 
Own children under 6 years only &   10.39  &  5.49  &   2.99  &   4.10  \\
Own children between 6 and 17 years only &  10.98  &  5.71  &  3.30  &  4.59  \\
Own chilren under 6 and between 6 and 17 years &  14.88  &  5.71  &  3.11  &  4.09  \\
No own children under 18 years &  12.84  &  6.05  &  4.03  &  5.20 \\
\hline
\end{tabular} \label{table:metrics_dhch_OwnChild}
\end{table}

\begin{table}[h!]
\caption{Production US 2020 DHCH: Householder Couple Type $\rho$-zCDP MAE}
\begin{tabular}{|R{7.1cm} |R{1.5cm}R{1.5cm}R{1.5cm}R{1.5cm}|}
\hline
Group & States & Counties & Places & Tracts \\
\hline 
Opposite-sex married couple household &   10.67  &   3.43  &   3.01  &   2.89  \\
Same-sex married couple household &  8.33  &  2.50  &  1.47  &  1.59  \\
Opposite-sex unmarried partner household &  10.00  &  3.39  &  2.38  &  2.46  \\
Same-sex unmarried partner households &  8.41  &  2.37  &  1.28  &  1.43  \\
\hline
\end{tabular} \label{table:metrics_dhch_coupleType}
\end{table}


\begin{table}[h!] 
\caption{Production US 2020 DHCH: Householder Race $\rho$-zCDP MAE}
\begin{tabular}{|R{7.1cm} |R{1.5cm}R{1.5cm}R{1.5cm}R{1.5cm}|}
\hline
Group & States & Counties & Places & Tracts \\
\hline 
White alone &   36.20  &   8.50  &   4.11  &   4.56  \\
Black/African American alone &  38.02  &  6.53  &  2.45  &  3.46  \\
American Indian/Alaska Native alone &  47.78  &  4.93  &  1.71  &  2.24  \\
Asian alone &  26.25  &  4.63  &  1.72  &  2.74  \\
Native Hawaiian/OPI* alone &  25.63  &  2.54  &  0.74  &  0.87  \\
Some Other Race alone &  20.16  &  5.36  &  2.28  &  3.21  \\
Two or More Races  &  21.92 &  5.75  &  2.80  &  3.71  \\
\hline
\multicolumn{5}{|p{14.5cm}|}{\footnotesize{*OPI stands for Other Pacific Islander.}}\\
\hline
\end{tabular} \label{table:metrics_dhch_hhrace}
\end{table}

\begin{table}[h!] 
\caption{Production US 2020 DHCH: Tenure by Householder Age $\rho$-zCDP MAE}
\begin{tabular}{|R{7.1cm} |R{1.5cm}R{1.5cm}R{1.5cm}R{1.5cm}|}
\hline
Group & States & Counties & Places & Tracts \\
\hline 
Owner occupied &  &  &  &  \\
\;\;\;Householder 15 to 24 years &  112.76  &  13.67  &  4.52  &  3.50  \\
\;\;\;Householder 25 to 34 years &  130.73  &  22.13  &  8.09  &  8.31  \\
\;\;\;Householder 35 to 54 years &  142.14  &  27.59  &  10.05  &  10.86  \\
\;\;\;Householder 55 to 64 years &  130.59  &  24.99  &  8.65  &  9.77  \\
\;\;\;Householder 65 years and over &  151.25  &  23.70  &  9.14  &  10.55  \\
Renter occupied &  &  &  &  \\
\;\;\;Householder 15 to 24 years &  83.04  &  15.06  &  5.56  &  5.91  \\
\;\;\;Householder 25 to 34 years &  108.04  &  19.25  &  6.65  &  7.83  \\
\;\;\;Householder 35 to 54 years &  79.10  &  18.85  &  7.87  &  8.97  \\
\;\;\;Householder 55 to 64 years &  107.25  &  17.48  &  5.61  &  6.45  \\
\;\;\;Householder 65 years and over &  94.39  &  14.10  &  5.10  &  6.58  \\
\hline
\end{tabular} \label{table:metrics_dhch_hhageXowned}
\end{table}

\begin{table}[h!] 
\caption{Production US 2020 DHCH: Tenure by Race $\rho$-zCDP MAE}
\begin{tabular}{|R{7.1cm} |R{1.5cm}R{1.5cm}R{1.5cm}R{1.5cm}|}
\hline
Group & States & Counties & Places & Tracts \\
\hline 
Owner occupied &   &   &   &   \\
\;\;\;White alone &  22.78  &  5.56  &  2.70  &  3.27  \\
\;\;\;Black/African American alone &  27.76  &  4.34  &  1.54  &  2.30  \\
\;\;\;American Indian/Alaska Native alone &  28.88  &  3.40  &  1.01  &  1.50  \\
\;\;\;Asian alone &  19.39  &  3.33  &  1.19  &  1.99  \\
\;\;\;Native Hawaiian/OPI* alone &  15.86  &  1.75  &  0.38  &  0.47  \\
\;\;\;Some Other Race alone &  13.33  &  3.81  &  1.60  &  2.24  \\
\;\;\;Two or More Races &  15.08  &  4.20  &  2.12  &  2.74  \\
Renter occupied &  &  &  &  \\
\;\;\;White alone &  17.08  &  4.70  &  2.26  &  2.95  \\
\;\;\;Black/African American alone &  13.92  &  3.94  &  1.42  &  2.26  \\
\;\;\;American Indian/Alaska Native alone &  20.80  &  2.83  &  0.99  &  1.47  \\
\;\;\;Asian alone &  12.10  &  2.67  &  0.95  &  1.63  \\
\;\;\;Native Hawaiian/OPI* alone &  13.47  &  1.59  &  0.42  &  0.61  \\
\;\;\;Some Other Race alone &  11.82  &  3.29  &  1.41  &  2.15  \\
\;\;\;Two or More Races &  12.24  &  3.77  &  1.66  &  2.49  \\
\hline
\multicolumn{5}{|p{14.5cm}|}{\footnotesize{*OPI stands for Other Pacific Islander.}}\\
\hline
\end{tabular} \label{table:metrics_dhch_hhraceXowned}
\end{table}

\begin{table}[h!] 
\caption{Production US 2020 DHCH: Household Size $\rho$-zCDP MAE}
\begin{tabular}{|R{7.1cm} |R{1.5cm}R{1.5cm}R{1.5cm}R{1.5cm}|}
\hline
Group & States & Counties & Places & Tracts \\
\hline 
1-person household &   18.10  &   18.83  &   8.07  &   7.92  \\
2-person household &  79.06  &  36.10  &  13.41  &  12.20  \\
3-person household &  65.67  &  29.59  &  10.55  &  11.67  \\
4-person household &  57.94  &  25.37  &  9.27  &  11.12  \\
5-person household &  51.45  &  24.91  &  9.10  &  9.35  \\
6-person household &  79.84  &  30.78  &  9.72  &  7.51  \\
7-or-more-person household  &  98.96  &  31.56  &  9.55  &  6.89  \\
\hline
\end{tabular} \label{table:metrics_dhch_hhsize}
\end{table}

\begin{table}[h!] 
\caption{Production US 2020 DHCP: Relationship to Householder $\rho$-zCDP MAE}
\begin{tabular}{|R{7.1cm} |R{1.5cm}R{1.5cm}R{1.5cm}R{1.5cm}|}
\hline
Group & States & Counties & Places & Tracts \\
\hline 
Householder &  23.02  &  6.70  &  4.32  &  4.26  \\
Opposite-sex husband/wife/spouse &  8.43  &  4.50  &  3.39  &  3.26  \\
Opposite-sex unmarried partner &  10.06  &  3.45  &  2.40  &  2.37  \\
Same-sex husband/wife/spouse &  8.39  &  2.58  &  1.44  &  1.47  \\
Same-sex unmarried partner &  6.92  &  2.33  &  1.37  &  1.35  \\
Biological son or daughter &  18.67  &  5.92  &  3.75  &  3.68  \\
Adopted son or daughter &  9.98  &  3.64  &  2.17  &  1.98  \\
Stepson or stepdaughter &  9.59  &  3.75  &  2.29  &  2.17  \\
Brother or sister &  8.35  &  3.30  &  2.15  &  2.18  \\
Father or mother &  7.02  &  3.08  &  2.08  &  2.06  \\
Grandchild &  11.35  &  3.91  &  2.60  &  2.58  \\
Parent-in-law &  6.25  &  2.63  &  1.47  &  1.57  \\
Son-in-law or daughter-in-law &  5.31  &  2.70  &  1.63  &  1.67  \\
Other relative &  11.02  &  3.93  &  2.33  &  2.33  \\
Roommate or housemate &  9.73  &  3.80  &  2.15  &  2.30  \\
Foster child &  6.61  &  2.58  &  1.46  &  1.23  \\
Other nonrelative &  10.80  &  3.84 &  2.37  &  2.21 \\
\hline
\end{tabular} \label{table:metrics_dhcp_relship}
\end{table}

\begin{table}[h!] 
\caption{Production US 2020 DHCP: 5 Year Age Groups $\rho$-zCDP MAE}
\begin{tabular}{|R{6.2cm} |R{1.7cm}R{1.7cm}R{1.7cm}R{2.0cm}|}
\hline
Group & Counties & Places & Tracts & AIR/ORT* \\
\hline 
Under 5 years &   9.52  &   4.47  &   5.28  &   3.95  \\
5 to 9 years &  10.64  &  4.36  &  5.84  &  4.35  \\
10 to 14 years &  10.63  &  4.50  &  5.95  &  4.61  \\
15 to 19 years &  8.31  &  4.52  &  4.39  &  2.84  \\
20 to 24 years &  4.57  &  2.78  &  3.06  &  2.08  \\
25 to 29 years &  4.59  &  2.80  &  3.09  &  2.17  \\
30 to 34 years &  4.39  &  2.77  &  3.02  &  2.12  \\
35 to 39 years &  6.56  &  3.01  &  4.06  &  3.29  \\
40 to 44 years &  6.36  &  2.94  &  4.00  &  2.86  \\
45 to 49 years &  6.25  &  2.93  &  3.95  &  2.89  \\
50 to 54 years &  6.22  &  2.99  &  3.96  &  3.20  \\
55 to 59 years &  6.17  &  3.07  &  3.96  &  3.35  \\
60 to 64 years &  8.08  &  3.80  &  4.99  &  3.33  \\
65 to 69 years &  7.90  &  3.63  &  4.75  &  3.53  \\
70 to 74 years &  6.01  &  2.84  &  3.67  &  2.83  \\
75 to 79 years &  5.87  &  2.71  &  3.50  &  2.76  \\
80 to 84 years &  5.88  &  2.55  &  3.33  &  2.54  \\
85 to 89 years &  5.39  &  2.29  &  3.10  &  1.90  \\
90 to 94 years &  5.29  &  2.19  &  2.75  &  1.55  \\
95 to 99 years &  4.77  &  1.69  &  2.00  &  0.99  \\
100 to 104 years &  3.14  &  0.94  &  0.77  &  0.38  \\
105 to 109 years &  1.29  &  0.27  &  0.14  &  0.40  \\
over 109 years &  0.64  &  0.03  &  0.05  &  0.19  \\
\hline
\multicolumn{5}{|p{14.5cm}|}{\footnotesize{*AIR/ORT stands for Federal American Indian Reservations/Off-Reservation Trust Lands.}}\\
\hline
\end{tabular} \label{table:metrics_dhcp_age}
\end{table}

Tables \ref{table:metrics_dhch_OwnChild} through \ref{table:metrics_dhch_hhsize} provide the $\rho$-zCDP MAE values of tabulations in the DHCH universe. As described previously, counts in this universe can be viewed as counts of households; alternatively, since there is one householder per household, counts in this universe can also be viewed as counts of householders. For the DHCP universe, Tables \ref{table:metrics_dhcp_relship} and \ref{table:metrics_dhcp_age} provide the $\rho$-zCDP MAE of counts of individuals grouped by relationship to the householder and by their age group, respectively.

\section{Conclusion and Progress after the 2020 Census DHC Release} \label{sec:nmfs}

As described in Section \ref{sec:intro}, one important advantage of formally private mechanisms like the DAS is that they are sufficiently transparent to allow users to construct confidence intervals of the confidential CEF-based tabulations in a straightforward manner. To describe how this can be done, more background on how the strategy matrices described in Section \ref{sec:accuracy_experiments} are used within the DAS will be helpful. Specifically, after spine optimization, the noisy measurements for each tabulation and each geographic unit are defined by adding discrete Gaussian noise to the CEF-based tabulation of the given geographic unit. If the PLB allocation of the geographic unit and tabulation pair is given by $\rho,$ the mean-zero noise added to each level of the CEF-based tabulation is given by a realization of a random variable from the discrete Gaussian distribution, $N_\zz(\mu=0, \sigma^2=1/\rho),$ which is defined so that its probability mass function is proportional to $\exp(-x^2/2 \sigma^2)$ for each $x\in \zz$ and zero otherwise \parencite{canonne2020discrete}. We call the collection of all such noisy measurements for a given DAS execution the noisy measurement files (NMFs). 

To support research and development on tools that would provide users inferences on underlying CEF-based tabulations, the Census Bureau has released NMFs for both the final 2010-CEF-based DDP production settings DAS implementations and the 2020 production DAS implementations as experimental data products. Specifically, Table \ref{tab:aws_open_data} provides AWS Open Data URLs for the DAS evaluation NMFs using our 2020 production settings, for both the Redistricting Data File and the DHC statistical data products and for both the 2020 production DAS executions and the DDPs DAS executions that used 2010 input data. Each of these four sites contains a README file that provides information on the format of the corresponding NMFs, an AWS Simple Storage Service (S3) path for the NMFs, and instructions on how to download these NMFs using the AWS S3 command line interface. 

\begin{table}[h!]
\caption{URLs for DAS NMFs hosted on AWS Open Data, for both the 2010 and 2020 input data vintages and for both the DHC and redistricting data file statistical data products} \label{tab:aws_open_data}
\begin{tabular}{lll}
Data Product                   & Vintage & URL  \\ \hline
\multirow{2}{*}{Redistricting} & 2010    & \url{https://registry.opendata.aws/census-2010-pl94-nmf/} \\
                               & 2020    & \url{https://registry.opendata.aws/census-2020-pl94-nmf/} \\ \hline
\multirow{2}{*}{DHC}           & 2010    & \url{https://registry.opendata.aws/census-2010-dhc-nmf/} \\
                               & 2020    & \url{https://registry.opendata.aws/census-2020-dhc-nmf/}\\ \hline
\end{tabular}
\end{table}

Subsets of the NMFs that provide query answers for queries that are at least as granular as the query of interest can be used to generate point estimates and confidence intervals of the confidential CEF-based tabulations via either simulation or a continuous Gaussian modeling assumption; see for example \parencite{nmfCIs, McCartan2023Making}. Alternatively, rather than using small subsets of the noisy measurements from the NMFs, \textcite{cumings2024full} provides a full-information confidence interval estimator, \textit{i.e.}, an estimator that uses all noisy measurements from the respective DAS execution. The Census Bureau is currently exploring releasing confidence intervals of CEF-based tabulations that are computed using an implementation of this approach as an experimental data product.

Confidence interval estimators based on the NMFs, such as the examples cited above, result in intervals that are not centered around the count implied by the MDF output from the DAS, and instead are centered around an unbiased estimate of the CEF-based tabulation. In contrast, \textcite{ashmead2024} provide several heuristic confidence interval estimators centered directly around the published tabulations that are based on MDFs generated by simulating the DHC DAS implementations, with the input of each simulation iterate being either the 2020 production DHC DAS output MDFs or DHC DAS output MDFs that used a 2010-vintage input CEF. By using MDFs that have already been protected by a formally private mechanism as input rather than the confidential CEF, this method avoids additional PLB expenditures. In some cases these confidence interval estimates are more narrow than those that are based on creating an unbiased estimator of the target query using the 2020 production DHC DAS NMFs. The Census Bureau has released the simulation MDFs required to compute these confidence interval estimates, and \textcite{ashmead2024} provide instructions for downloading these simulation MDFs.

\printbibliography
\end{document}